\documentclass[twocolumn,showpacs,preprintnumbers,amsmath,amssymb,groupedaddress,prl]{revtex4}

\usepackage{graphicx}
\usepackage{dcolumn}
\usepackage{bm}
\usepackage{epsfig}

\def\e{\begin{equation}}
\def\f{\end{equation}}
\def\=#1{\overline{\overline #1}}

\def\-#1{{\bf #1}}
\def\.{\cdot}

\def\E{\epsilon}

\begin{document}

\title{Experimental demonstration of sub-wavelength image channeling using\\
a capacitively loaded wire medium}

\author{Pekka Ikonen$^1$, Pavel Belov$^{2,3}$, Constantin Simovski$^{1,3}$, and Stanislav
Maslovski$^1$}

\affiliation{$^1$Radio Laboratory/SMARAD, Helsinki University of
Technology, P.O.~Box 3000, FI-02015 TKK, Finland\\
$^2$Queen Mary College, University of London, Mile End
Road, London, E1 4NS, United Kingdom\\
$^3$Photonics and Optoinformatics Department, St.~Petersburg State
University of Information Technologies, Mechanics and Optics,
Sablinskaya 14, 197101, St.~Petersburg, Russia}

\begin{abstract}
In this letter we experimentally demonstrate a possibility to
achieve significant sub-wavelength resolution of a near-field image
channeled through a layer of an electromagnetic crystal. An image
having radius of $\lambda/10$ has been realized using an
electrically dense lattice of capacitively loaded wires. The loading
allows to reduce the lattice period dramatically so that it is only
a small fraction of the free-space wavelength. It is shown that
losses in the structure only decrease the total amplitude of the
image, but do not influence the resolution.
\end{abstract}


\maketitle

It was theoretically shown by Notomi \cite{NotomiPRB} and
experimentally proven by Parimi et.~al \cite{Parimilens} that in
photonic crystals (PC) or electromagnetic crystals (EC) negative
refraction can be achieved at frequencies close to the
electromagnetic band-gap edges. Developing this direction of
studies, Luo et.~al. introduced a flat \emph{superlens} formed by a
slab of PC in \cite{Allanglediag}, and studied theoretically the
possibility for sub-wavelength imaging in a layer of EC
\cite{Allanglelight}. Recently, it has been shown that with PC/EC
based structures neither negative refraction or surface plasmon
excitation is required to obtain a point source image whose size is
smaller than $\lambda/2$. In this case the subwavelength image size
has theoretically been achieved when the crystal iso-frequency
contour has a special (flat) shape \cite{Li,Chien,Kuo,Zhang1,Belov}.
In this letter we experimentally validate the recently proposed
concept of \emph{sub-wavelength image channeling} \cite{Belov} that
explains the aforementioned effect by a transformation of evanescent
modes into propagating crystal eigenmodes, and describes a special
regime for the wave propagation. This regime has been called in the
literature as {\it self collimation} \cite{Li}, {\it directed
diffraction} \cite{Chien}, {\it tunneling} \cite{Kuo}, {\it absolute
negative refraction} \cite{Zhang1}, and {\it self-guiding}
\cite{Chigrin}. In subwavelength image channeling the crystal slab
effectively operates as a transmission device transferring the near
field distribution of a source from the front interface to the back
interface. It is conceptually important to note that the slab is not
a lens in the optical point of view, and that no focusing is
incorporated to the phenomenon. When the slab operates within the
channeling regime and the source is located close to the front
interface, the evanescent free-space harmonics transform into
propagating crystal eigenmodes at the first crystal interface
\cite{Belov}. The near field distribution of the source is carried
by these eigenmodes to the back interface where it diffracts into
the free space. Very close to the back interface the channeled
distribution can be detected.

\begin{figure}[b!]
\centering \epsfig{file=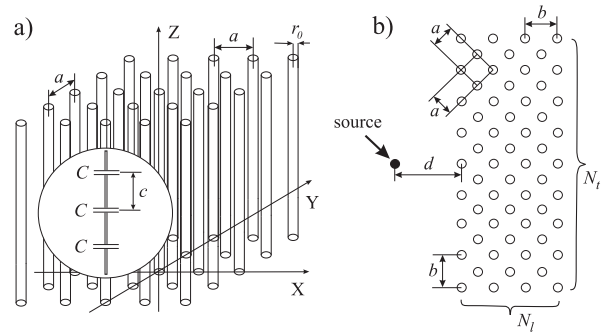, width=7.8cm} \caption{a)
Geometry of capacitively loaded wire medium (CLWM). b) Geometry of
the crystal sample.} \label{geom}
\end{figure}

The crystal structures used to theoretically demonstrate the
sub-wavelength image channeling include lattices of dielectric rods
\cite{Li,Kuo} and lattices of metal rods coated with
high-permittivity dielectrics \cite{Zhang1}. We propose to utilize
\emph{capacitively loaded wire medium} (CLWM) \cite{lwPRE} for
implementing the operational regime, see Fig.~\ref{geom}.a. Compared
to conventional wire medium \cite{Nicorovici} or lattices of
dielectric rods, CLWM offers additional features which make it
superior in design. Authors of \cite{lwPRE} have shown that CLWM has
a band-gap near the resonant frequency of a single capacitively
loaded wire (the so called \emph{resonant band-gap}). By changing
the value of the load capacitance, the location of this band-gap can
be conveniently tuned. Moreover, high value of capacitance allows to
locate the resonant band-gap significantly below the first lattice
resonance, thus the needed regime can be achieved with an
electrically dense lattice. Obviously, the image size is restricted
by the lattice period, thus, the more (electrically) dense is the
lattice, the smaller can be the image size (compared to the
wavelength in free space).

\begin{figure}[b!]
\centering \epsfig{file=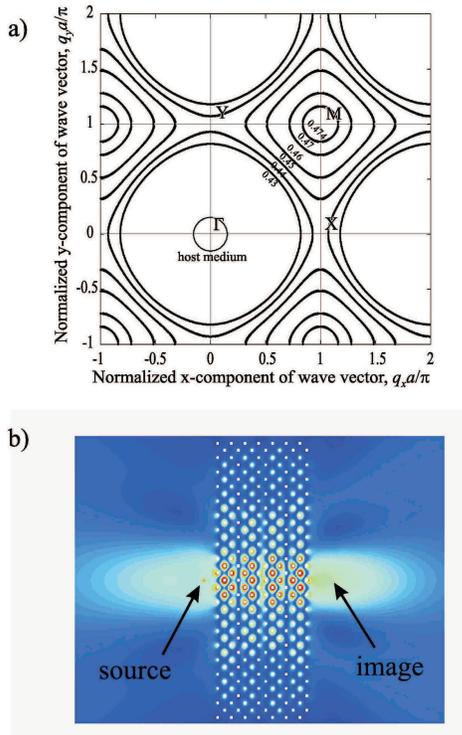, width=6.2cm} \caption{a)
Isofrequency contours for CLWM. The numbers correspond to values of
normalized frequency $ka$. b) Simulated intensity distribution.}
\label{izo}
\end{figure}

\begin{figure}[ht!]
\centering \epsfig{file=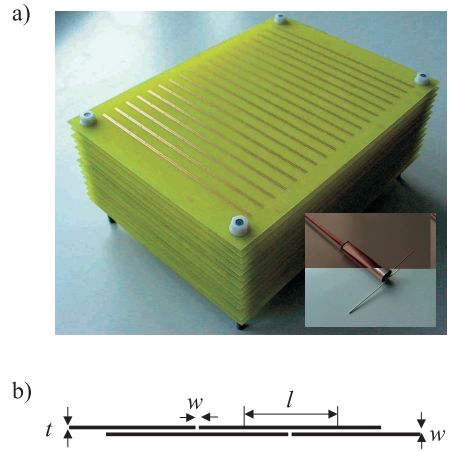, width=5.8cm} \caption{a) The
implemented CLWM crystal sample and the probe used in the
measurements. b) A schematic illustration of the loaded wire (piece
of it).} \label{photo_lens}
\end{figure}

For implementing the proposed regime we use the following geometry
and structural dimensions (see Fig.~\ref{geom}): The lattice period
$a=10$ mm, the radius of wires $r_0=0.058a$, the insertion period
for the loads $c=2.2a$, and loading capacitance $C_0=0.5$ pF. The
crystal sample has dimensions $N_l=14$ and $N_t=21$, where $N_l$ is
the number of rows of wires, and $N_t$ is the number of columns of
wires. $N_l=14$ is chosen to meet Fabry-P\'erot resonance (FPR)
condition at the normalized frequency $ka=0.46$ \cite{Belov}, where
$k$ is the wave number in free space ($a/\lambda=0.073$).
Fig.~\ref{izo}.a shows isofrequency contours for the frequency
region near the lower edge of the resonant band-gap. The
isofrequency contour of the host matrix for $ka=0.46$ is shown as
the small circle around $\Gamma$ point. The part of the isofrequency
contour of the crystal corresponding to $ka=0.46$, and located
within the first Brillouin zone is practically flat. This part is
perpendicular to the diagonal of the first Brillouin zone. Thus, in
order to achieve channeling regime we orient interfaces of the slab
orthogonally to (11)--direction of the crystal as shown in
Fig.~\ref{geom}.b. Fig.~\ref{izo}.b depicts the simulated intensity
distribution when using a line source as excitation \cite{Belov}.
There is a clear channel through the slab indicating the propagation
of the excited crystal eigenmodes. A bright spot having a radius of
$\lambda/6$ (determined from the intensity distribution at level
max(intensity)/2) is seen in the image area.

Fig.~\ref{photo_lens}.a shows a photograph of the implemented
CLWM sample. To ease the fabrication, the capacitively loaded wires
are implemented as printed strips on top of FR4--substrate (relative permittivity $\E_{\rm
r} \sim 4.5(1 - j0.01)$, thickness of the plates is 1.0 mm).
According to numerical simulations, losses introduced by the rather
low quality substrate are within an acceptable range. Thus, despite
the lowered transmission level at the desired FPR, the achieved
intensity distribution should be only moderately disturbed. To
implement the desired distributed capacitance level we utilize the
structure shown in Fig.~\ref{photo_lens}.b. Since the transverse
thickness of the structure (strip) is very small, the incoming wave
sees the structure as a continuous $C-$loaded strip. Also, in this
 structure the capacitance is well distributed along the
strip leading to highly uniform current distribution. Estimated
dimensions for the structure to produce the desired distributed
capacitance level of 11 fF$\cdot$m are $l=$ 22.0 mm, $t=$ 0.575 mm,
and $w=$ 1.1 mm (see Fig.~\ref{photo_lens}.b).

\begin{figure}[t!]
\centering \epsfig{file=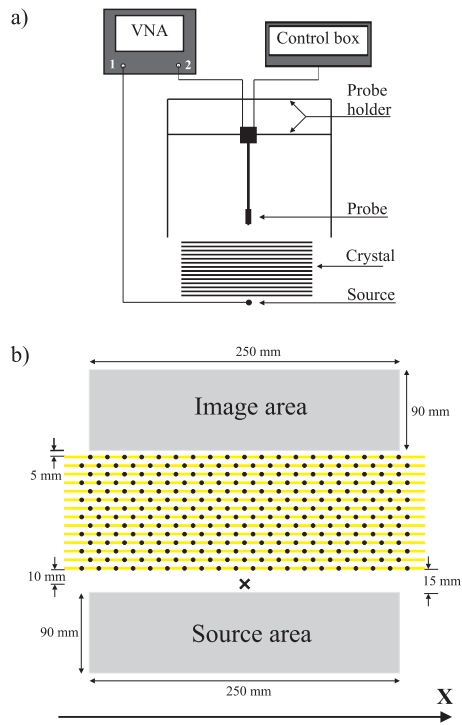, width=6.0cm} \caption{a) A
schematic illustration of the measurement setup. b) The measurement
area. The black dots denote the loaded strips, the cross denotes the
source.} \label{meas_path}
\end{figure}

The measurement setup is schematically depicted in
Fig.~\ref{meas_path}.a. The source and the probe used to scan the
field distribution are connected to a vector network analyzer (VNA).
The probe holder is connected to a control box responsible for the
movement of the probe. VNA contains an in-house program for
recording electric field values. To maximize the sensitivity of the
measurement setup, the source and the probe are both based on
resonant dipoles operating at the measured FPR (see
Fig.~\ref{photo_lens}.a for a photograph of the probe).
Fig.~\ref{meas_path}.b shows the measurement areas, and the
orientation of the CLWM crystal during the measurements. The spatial step of
the probe is 5 mm.

\begin{figure}[hb!]
\centering \epsfig{file=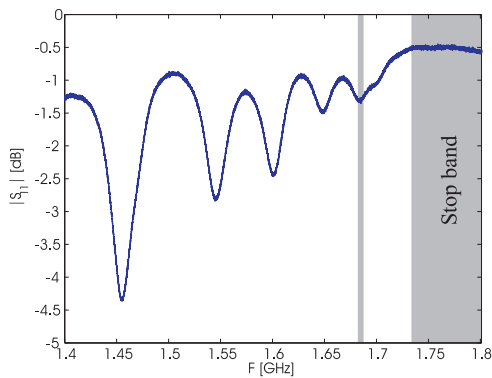, width=6.5cm} \caption{Measured
$S_{11}-$level of the slab. The desired Fabry-P{\'e}rot resonance is
shown with a vertical line.} \label{s11}
\end{figure}

First, we need to locate the desired FPR for which the slab has been
tuned. This frequency lies approximately at the lower edge of the
first crystal stop band. To locate the FPR we measure the reflection
level from the slab using a wide-band horn. The result is presented
in Fig.~\ref{s11}. Undoubtedly, there are strong near field
interactions between the slab and the horn so Fig.~\ref{s11} does
not properly show the true reflection properties of the slab.
Nevertheless, the stop band edge at 1.73 GHz, and the FPR at 1.684
GHz are seen in Fig.~\ref{s11}: only a traveling wave is capable of
producing the resonant peaks shown in Fig.~\ref{s11}. Inside the
stop-band of the structure a wave decays exponentially and the
amplitude of the reflection coefficient approaches unity. Thus, when
the resonant peaks are not anymore seen and the reflection
coefficient amplitude is close to unity we have found the stop-band
edge.

The dielectric plates have a visible influence on the transmission
characteristics: the lower edge of the resonant stop band has
shifted from the predicted (for wires in free space) 2.20 GHz
\cite{Belov} to 1.73 GHz. Consequently, there is most likely some
discrepancy in the estimated parameters used to predict the
capacitance level. From the measured location of the lower stop band
edge we estimate that effectively the distributed capacitance level
at the FPR is 17.6 fF$\cdot$m. According to the numerical
simulations higher capacitance level should improve the resolution.

\begin{figure}[b!] \centering \epsfig{file=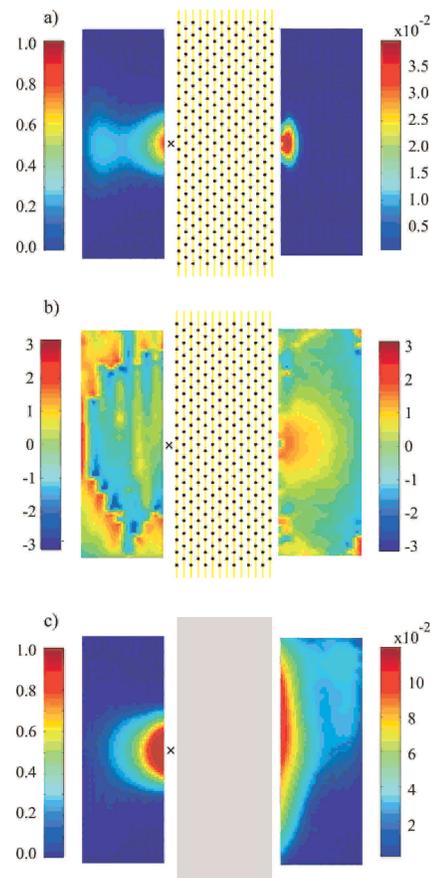, width=5.5cm}
\caption{a) The measured intensity distribution (in arb.~units). b)
The measured phase distribution of electric field (in radians). c)
The measured intensity distribution without the CLWM crystal (in
arb.~units). In a) and c) the intensity values in the image area
have not been normalized to better see the details of the intensity
distribution.} \label{intensity}
\end{figure}

The following measurements have been performed at frequency 1.684
GHz which corresponds to the FPR identified from the reflection
measurement. Fig.~\ref{intensity}.a shows the measured intensity
distribution in the source and the image area. We can observe that
there is a clear and symmetrical image at the predicted image
location. The radius of the image (determined at level
max(intensity)/2) is approximately $0.10\lambda$, thus we are
dealing with a significant sub-wavelength resolution. It is worth
noticing that due to the nature of the measurement probe, the probe
is inherently incapable of averaging the detected transmitted field
(as is the case e.g.~with loop detectors \cite{Marqueslatest}).
Since the dipole arm is very thin, the exact distribution of the
transmitted field can be very accurately measured, and the
resolution of the image hardly depends on the electrical size of the
probe. Contrary to the simulations, we can observe from
Fig.~\ref{intensity}.a that as the probe moves closer to the slab
interface, the field strength slightly decreases. This is most
likely caused by destructive interaction between the metal parts of
the probe and the metal strips leading to the loss of the resonant
condition in the vicinity of the wire surface (similar effect has
been observed with loop shaped metal particles and loop shaped metal
detector \cite{Marqueslatest}).

The measured phase distribution of the electric field is presented
in Fig.~\ref{intensity}.b. In the image area we can identify the
phase front of a wave diverging from the image. In the source area
the phase distribution is disturbed by reflections from the slab.
Fig.~\ref{intensity}.c shows the measured intensity distribution
without the crystal (the shadowed region shows the space previously
occupied by the slab). Slight asymmetry is seen in the cylindrical
intensity profile. This weak asymmetry is expected to be caused by
parasitic currents induced to the balun of the source probe (the
transversal thickness of the balun is not negligible). The effect of
parasitic radiation is not seen in the source area when the field
level produced by the dipole radiation is strong. Moreover, the
parasitic radiation does not destroy the symmetry of the intensity
profile when the slab is present. This is due to the fact that
outside the channel the field level is very low (most of the energy
flows inside the channel), thus the parasitic radiation directed
away from the symmetry axis of the slab attenuates rapidly before
approaching the image area.

\begin{figure}[t!]
\centering \epsfig{file=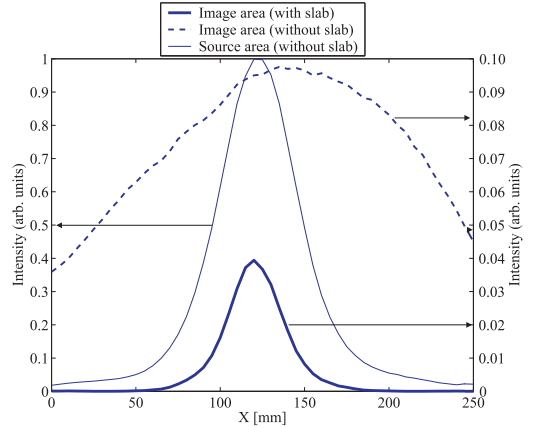, width=6.8cm} \caption{Measured
intensity distribution along the axis X (parallel to the slab
interface). The arrow indicates the scale which is used to read the
intensity values. Intensity in the image area has been measured at
10 mm distance from the slab interface. Intensity in the source area
has been measured at 10 mm distance from the source.}
\label{close_look}
\end{figure}

The measured transversal intensity distributions are shown in
Fig.~\ref{close_look} (see Fig.~\ref{meas_path}.b for definition of
the coordinate system). 
The measured maximum field value is approximately 35 percent lower
when using the crystal slab than in the free space measurement. The
biggest reason for this degradation is the strong reflection
occurring at the first crystal interface. The lossy dielectric
plates also degrade the amplitude level of the channeled field. It
is important to note, however, that the image shape is not affected
by the losses. This important practical property is inherent to the
subwavelength image channeling: the propagating waves in a lossy
material experience the same decay irrespective of their transversal
wave vector components which are responsible for the shape of the
image.

In the present letter we have experimentally validated the concept
of sub-wavelength image channeling using a slab of an
electromagnetic crystal. It has been shown that losses influence
only to the total amplitude of the image, but do not disturb its
shape. We have shown that an image with radius of $\lambda/10$ can
be achieved using an electrically dense lattice of capacitively
loaded wires. The image size is noticeably below the theoretical
diffraction limit, and it is twice smaller than the minimum image
size practically obtained in the modern high-precision microscopy.
Compared to the results obtained in the optical regime \cite{Fang}
the quality of imaging is improved twice. In the microwave regime
the proposed structure could be used as a wave channel multiplexer
or a decoupler for antenna systems.

Useful discussions with Prof.~Sergei Treyakov are acknowledged. One
of the authors (P.I.) wishes to thank Mr.~Eino Kahra from the Radio
Laboratory TKK for assistance in manufacturing the prototype.


\bibliography{CLWM_superlens}
\end{document}